\documentstyle[aps,prl,preprint]{revtex}

\begin{document}

\draft

\title{
Quantum Reservoir Engineering
}

\author{J.F. Poyatos$^*$, J.I. Cirac,$^*$ and P. Zoller}

\address{
Institut f\"ur Theoretische Physik, Universit\"at Innsbruck,
Technikerstrasse 25, A--6020 Innsbruck, Austria.
}

\date{\today}

\maketitle

\begin{abstract}
We show how to design different couplings between a single ion
trapped in a harmonic potential and an environment. This will
provide the basis for the experimental study of the process of
decoherence in a quantum system. The coupling is due to the
absorption of a laser photon and subsequent spontaneous
emission. The variation of the laser frequencies and
intensities allows one to ``engineer'' the coupling and select
the master equation describing the motion of the ion.
\end{abstract}


\narrowtext
\newpage

One of the fundamental questions of Physics is understanding
the borderline between microscopic phenomena ruled by quantum
mechanics and the macroscopic world of classical physics. In
particular, according to quantum mechanics \cite{Di84}, a
system can exist in a superposition of distinct states, whereas
these superpositions seem not to appear in the macroscopic
world. One possible explanation of this paradox \cite{Zu91} is
based on the fact that systems are never completely isolated
but interact with the surrounding environment, that contains a
large number of degrees of freedom. The environment influences
the system evolution which continuously decoheres and
transforms system superpositions into statistical mixtures
which behave classically \cite{Zu91,Wa85}. This
problem is directly related to the problem of measurement in
quantum theory \cite{Ne32,Wh83} where the system to
be measured is described by quantum mechanics and the
measurement apparatus is assumed to behave classically. Apart
from this fundamental point of view a more practical aspect is
the question to what extent one can preserve quantum
superpositions, which is the basis of potential applications of
quantum mechanics, such as quantum cryptography and computation
\cite{Di95,Ch95}.

From the theoretical point of view, quantum decoherence has
been studied extensively \cite{Zu91,Wa85,Da93,Ga94,Go95,Ga92}.
Most of the effort has been focused on the decoherence of a
harmonic oscillator (the system) due to the coupling to a
reservoir consisting of oscillators (the environment).
According to these studies, the decoherence process depends
critically on the form of the coupling between the system and
the environment. On the experimental side, however, there have
not been any systematic investigations. This is due to the lack
of experimentally accessible systems where both the time scale
and the form of the system--environment coupling can be changed
in a controlled way.

In this paper we will show how to ``engineer'' the
system--environment coupling in a situation that is
experimentally accessible with existing technology. The system
of interest will be an ion confined in a electromagnetic trap,
and the environment will be the vacuum modes of the
electromagnetic field. This corresponds to an experimental
realization of a harmonic oscillator coupled to a reservoir of
oscillators. The coupling between our system and the
environment takes place through the recoil experienced by the
ion when it interchanges photons with the electromagnetic
field. As we will show below, this coupling can be manipulated
by laser radiation. Variations of the laser frequency and
intensity allows one to engineer such a coupling.

Laser cooled trapped ions \cite{Wi87} are a unique
experimental system: unwanted dissipation can be made
negligible for very long times, much longer than typical times
in which an experiment takes place. Furthermore, arbitrary
quantum states of the ion's motion can be synthesized and
coherently manipulated using laser radiation \cite{Ci93}. In
addition, the state of motion can be completely determined in
the sense of tomographic measurements \cite{Po95}.  In a
series of remarkable experiments, Wineland and collaborators
have generated a variety of non--classical states of ion motion
\cite{Mo95b,Mo96}. In particular, they have been able
to produce \cite{Mo96} a so--called ``Schr\"odinger cat state''
\cite{Sc35} corresponding to
\begin{equation}\label{eq:cat}
|\Psi\rangle \propto | \alpha\rangle + |- \alpha \rangle,
\end{equation}
with $|\alpha\rangle= \sum_{n=0}^\infty \alpha^n /\sqrt{n!}
|n\rangle$ a coherent (quasiclassical) state. In Fig.~1(a) we
have plotted the density operator for such a state in the
position representation, i.e.~(the real part of)
$\rho(x,x')=\langle x| \rho | x' \rangle$. The peaks near the
diagonal correspond to two possible localizations of the
particle, whereas the other two peaks are related to the
coherences that are responsible from the quantum behavior
\cite{Zu91}.

The interaction of a Schr\"odinger cat state with the
environment has been the paradigm of decoherence of
superposition states. As first argued by Zurek \cite{Zu91} (see
also Refs.~\cite{Wa85,Zu91,Da93,Ga94}), for a
coupling which is linear in the system coordinates, a
macroscopic superposition of the form (\ref{eq:cat}) decays to
a statistical mixture $
\rho \propto |\alpha \rangle \langle \alpha | +
|-\alpha \rangle \langle -\alpha |$,
on a short time scale (decoherence time) which is related to
the size of the cat ($|\alpha|^2$) and is much faster than the
energy dissipation time: this provides an explanation for the
absence of superpositions in the macroscopic world \cite{Zu91}.
We emphasize that the decoherence process of (\ref{eq:cat}) is
sensitive to the form of the reservoir coupling.  For some
quadratic couplings, for example, the decoherence and energy
dissipation time can become identical \cite{Ga94};
moreover, there exist interactions which allow Schr\"odinger
cat states to be stable, and, what is more surprising,
dissipation can drive a system into a steady state of the form
(\ref{eq:cat}) \cite{Ga94}. For example, in Fig.~1(b,c) the
decay of a Schr\"odinger cat under linear and quadratic
coupling is illustrated: for a linear coupling (Fig.~1 b) the
nondiagonal peaks (coherences) of the density matrix decay much
faster than for the quadratic couplings (Fig.~1 c). We will
show that all these theoretical predictions can be tested
experimentally for the case of a trapped ion.

The process of decoherence can be analyzed in detail under very
general assumptions invoking to the so--called Markov approximation,
which considers the correlation time for the environment to be much
shorter that the evolution time of the system due to the coupling
\cite{Ga92}.  In this case the interaction of a system with an
environment is described in terms of a master equation. For a single
decoherence channel this equation has form ($\hbar=1$)
\begin{equation}
\label{MEgen0}
\dot \rho = \gamma
(2f\rho f^\dagger - f^\dagger f\rho - \rho f^\dagger f),
\end{equation}
Here $\rho$ is the reduced density operator for the system in
the interaction picture after tracing over the reservoir. The
operator $f$ and the parameter $\gamma$ reflects the
system--environment coupling. For a harmonic oscillator $f$
will be a function of the creation and annihilation operators
$a$ and $a^\dagger$, which are defined as usual
$X = 1/(2M\nu)^{1/2} (a^\dagger + a)$, $
P = i (M\nu/2)^{1/2} (a^\dagger -a)$
where $X$ and $P$ are the position and momentum operators and
$M$ the particle's mass.  According to Zurek \cite{Zu91}, the
coupling with the environment singles out in a quantum system a
preferred set of states, sometimes called ``the pointer
basis''. This basis depends on the form of the coupling $f$.
For example, for $f = X$ the pointer basis is the position
eigenstates. The density operator describing the system evolves
in such a way that it rapidly becomes diagonal in this
preferred basis, which is usually connected to the
disappearance of quantum interferences. Our goal is now to find
an experimental realization of the master equation
(\ref{MEgen0}) for different system--reservoir couplings
$f\equiv f(a,a^\dagger)$.

Let us consider a single ion moving in a one--dimensional harmonic
potential. The ion interacts with a laser in a standing wave
configuration of frequency $\omega_L$, close to the transition
frequency $\omega_0$ of two internal levels $|g\rangle$ and
$|e\rangle$.  Using standard methods in quantum optics based on the
the dipole, Born--Markov, and rotating wave approximations, the master
equation that describes this situation can be written in the general
form
\begin{equation}
\label{MEgen2}
\dot \rho = -i H_{\rm eff} \rho + i \rho H_{\rm eff} ^\dagger
+ {\cal J} \rho,
\end{equation}
with
\begin{mathletters}
\begin{eqnarray}
H_{eff} &=& H_{tp} + H_{int} + H_{cou}
- i \frac{\Gamma}{2} |e\rangle\langle e|,\\
\label{Jop}
{\cal J} \rho &=& \Gamma  \int_{-1}^{1} du N(u)
e^{-i\eta u (a+a^\dagger)} \sigma_- \rho \sigma_+
e^{i\eta u (a+a^\dagger)}
\end{eqnarray}
\end{mathletters}
where $H_{tp} = \nu a^\dagger a$,
$H_{int} = \frac{1}{2} \omega_0 \sigma_z$, and
$H_{cou} = \frac{\Omega}{2} \sin[\eta(a+a^\dagger)+\phi]
(\sigma_+ e^{-i\omega_L t} + \sigma_- e^{i\omega_L t})$,
give the free Hamiltonian for the motion in the trap, the
internal two--level system Hamiltonian, and the one describing
the coupling with the lasers, respectively. Here,
$\sigma_+=|e\rangle\langle g|=(\sigma_-)^\dagger$ and
$\sigma_z=|e\rangle\langle e|- |g\rangle\langle g|$ are usual
spin--$\frac{1}{2}$ operators describing the internal
transition, $\nu$ is the trap frequency, $\Gamma$ the
spontaneous emission rate, and $\eta=(k_L^2/2M\nu)^{(1/2)}$ the
Lamb--Dicke parameter which is the ratio of the recoil
frequency $k_L^2/2M$ to the trap oscillation frequency $\nu$.
In the expression for the superoperator $\cal J$, the
exponentials are related to the photon recoil that takes place
in each spontaneous emission process, and the integral takes
into account the different angles at which that photon can be
emitted, with a normalized dipole pattern $N(u)$. In the
Hamiltonian describing the coupling with the lasers, $\Omega$
is the Rabi frequency, and $\phi$ characterizes the relative
position of the trap center with respect to the node of the
laser standing wave.  Here we will assume that either $\phi=0$
(excitation at the node of the standing wave) or $\phi=\pi/2$
(excitation at the antinode).

We will proceed now by simplifying the master equation for the
ion in a regime defined by three limits which are typically
fulfilled in experiments \cite{Mo95b,Mo96}: (i)
Lamb--Dicke, (ii) strong confinement (iii) low intensity.  The
first one allows to expand the above master equation in terms
of the Lamb--Dicke parameter $\eta\ll 1$, retaining only the
orders that contribute to the dynamics. The second one assumes
$\Gamma \ll \nu$ and together with the third one allows to
include in the coupling Hamiltonian only on--resonance terms
(secular approximation). Finally, the third one assumes a
sufficiently low laser intensity (the specific form of this
limit will be given later), and will serve us to adiabatically
eliminate the internal excited level $|e\rangle$.

Let us start by simplifying the coupling Hamiltonian under the
above limits. To do that, we move to a rotating frame defined
by the unitary operator ${\cal U}=e^{-i(H_{tp}+H_{int})t}$.
Following Ref.~\cite{Ci93} we assume that: (i) For excitation
at the node ($\phi=0$), $\delta=\omega_L-\omega_0=(2k+1)\nu$
($k=0,\pm 1,\ldots$) (ii) For excitation at the antinode
($\phi=\pi/2$), $\delta=2k\nu$ ($k=0,\pm 1,\ldots$). In this
rotating frame, after performing the rotating wave
approximation and the Lamb--Dicke expansion, we obtain
$H_{cou} = \frac{\Omega'}{2} (\sigma_+ f + f^\dagger \sigma_-)$,
where both $\Omega'$ and the form of the operator $f$ depend on
the frequency of the laser. For example, for $\delta=-\nu$, we
have $f= a$, and $\Omega'=\Omega\eta/2$, whereas for
$\delta=-2\nu$, $f=a^2$, and $\Omega'=-\Omega\eta^2/6$. Apart
from the strong confinement, in the first case, the secular
approximation can be performed for $\Omega' \ll \nu$, whereas
in the second case it is needed $\Omega^2/\nu \ll \Omega'$.
This two conditions can always be fulfilled for low enough
laser intensity, and together with $\Omega'\ll \Gamma$ define
the low intensity limit.

In the next step we eliminate adiabatically the internal
excited state using standard procedures of quantum optics
\cite{Ga92}. Physically, since $\Omega'\ll \Gamma$ the ion
practically spends no time in the excited level and therefore
we can eliminate it. Finally, expanding in powers of $\eta$ we
find the desired master equation (\ref{MEgen0}), with
corrections of the order $\eta^2$. The master equation will be
valid for times such that these corrections are not important,
that is for times $t \ll (\gamma \eta ^2\bar n)^{-1}$ where
$\bar n$ is the typical phonon number of the state of the ion.
Nevertheless in the Lamb--Dicke limit this time can be much
longer than the time required to reach the steady state using
the approximated master equation. Note that through the
adiabatic elimination we are coupling effectively the motion of
the ion with the environment.  The fact that this coupling
takes place through the absorption of laser photons, and we
have to choose the way on how this actually happens, allows one
to manipulate the coupling system--environment.

According to our analysis, by varying the laser frequency we
obtain the master equation (\ref{MEgen0}) with different
coupling operators $f$. In Fig.~2 we have illustrated the laser
configurations which produce several $f$ operators. In
Fig.~2(a), for example, the laser is tuned to the so-called
``lower motional sideband,'' $\delta = -\nu$, and the ion is
located at the node of the standing wave field which leads to a
coupling operator $f = a$. This can be easily understood by
noting that in each absorption and spontaneous emission cycle
one phonon is annihilated on a time scale given by the optical
pumping time.  Similarly, in Fig.~2(b) the laser is tuned to
the ``second lower sideband'' $\delta = -2\nu$ at the antinode
of the laser standing wave which gives the two--phonon coupling
$f = a^2$.  These two cases of linear and quadratic coupling
correspond to the two examples discussed in Figs.~1(b,c). In
fact, these figures were obtained by a numerical solution of
the full master equation (\ref{MEgen2}) with quantum
Monte--Carlo wavefunctions simulations \cite{Zo95}.
As noted before, the
decoherence acts in a different way depending on the coupling
operator, according to our previous discussion.

It is simple to generalize the above derivation to find
situations with other interesting (and perhaps unusual)
coupling operators $f$. For example, consider the case in which
two lasers of frequency $\omega_0 + \nu$ and $\omega_0-\nu$
interact with the ion [Fig~2(c)]. This corresponds to a
coherent excitation of the lower and upper motional sidebands
\cite{Ci93}. In this case, following the same arguments, one
can easily show that the operator is $f=\mu a + \nu a^\dagger$,
where $\mu^2-\nu^2=1$ and $\mu/\nu$ is the quotient of the Rabi
frequencies. This operator corresponds to a squeezed vacuum
coupling which has been the basis for numerous theoretical
predictions in quantum optics \cite{Ga92}. In particular,
choosing equal Rabi frequencies, the coupling is
$f=a+a^\dagger\propto X$. This corresponds to the case analyzed
theoretically by Unruh and Zurek, Caldeira and
Legget, and other authors \cite{Zu91,Wa85} to
describe the decoherence process in terms of the projection of
the state of the system onto the pointer basis given, in this
case, by the position eigenstates.  Another interesting
combination of lasers [Fig.~2(d)] yields
$f=(a-\alpha)(a-\beta)$, where $\alpha$ and $\beta$ are given
complex numbers.  For $\alpha=-\beta$ the Schr\"odinger cat
state (\ref{eq:cat}) is an eigenstate of this operator with
zero eigenvalue, and thus this state does not decohere under
this form of coupling. As an aside we note that one can employ
this particular form of system--reservoir coupling to generate
a cat state (\ref{eq:cat}) by choosing as the initial state the
ground level $|0\rangle$ \cite{Ga94,Ma96}. Tuning a laser
on resonance at the antinode of a standing light wave one can
design the coupling in the form of a quantum nondemonolition
measurement of the phonon number, $f=a^\dagger a$, with the
Fock states as the pointer basis. Using more
complicated laser configurations one can readily show that, for
example, $f$--operators like $a(a^\dagger a - n)$ can be
engineered. The operator $a^\dagger a - n$ projects out a
subspace with exactly $n$ phonons. Finally, exotic combination
of operators such as $a(a^\dagger a - n) - a^\dagger(a^\dagger
a - n-1)) -1$ can be realized, such that  the system is driven
into the superpositon state $|n\rangle+|n+1\rangle$.

Obviously, there are numerous possibilities to generalize the
concept of reservoir engineering in ion traps. First of all,
decoherence of a two or three--mode system can be studied by
considering the two or three dimensional motion of a trapped
ion, respectively. Furthermore, a master equation with more
than one decoherence channel, i.e., an equation containing sums
of damping terms of the form (\ref{MEgen0}) with different
operators $f_i$ ($i=1,\ldots,N$) \cite{Ga92}, can also be
easily implemented . This can be accomplished by exciting
transitions with several incoherent lasers. Another important
generalization concerns the possibility of coupling  a
two--level system to a harmonic oscillator (Jaynes--Cummings
model) which in turn is coupled to an environement. In
particular, this will allow to test experimentally one of the
outstanding predictions of quantum optics \cite{Ga86}, namely
the damping of a two--level system interacting with a squeezed
reservoir. Finally, these ideas can be extended to
linear ion traps \cite{Ci95} in order to study collective
effects in an $N$--atom $+$ harmonic oscillator system.

In summary, we have shown how the coupling of a harmonic
oscillator (represented by the motion of a trapped ion) to an
environment can be engineered. We believe that this opens a new
field in the sense that it will allow for the first time to
study experimentally in a controlled and systematic way the
effects of decoherence in a quantum system.




\begin{figure}
\caption{(a) $\rho(x,x')$ for the state (\ref{eq:cat})
($\alpha=3$). (b,c)
Numerical simulation of the interaction with a laser for a
time $\tau=0.06\gamma^{-1}$ and $\eta=0.03$: (b)
$\omega_L=\omega_0-\nu$, ($f=a$); (c) $\omega_L=\omega_0-2\nu$,
($f=a^2$).
}
\end{figure}

\begin{figure}
\caption{
Laser configurations for several coupling operators $f$. (a) Laser
tuned to $|n,g\rangle \rightarrow
|n-1,e\rangle$, which rapidly decays into the state
$|n-1,g\rangle$ leading to $f=a$. (b) $f=a^2$. (c) $f=\mu a +
\nu a^\dagger$.
(d) $f=(a-\alpha)(a-\beta)$.
}
\end{figure}


\begin{references}

\bibitem[*]{JIC}
Permanent address: Departamento de Fisica, Universidad de
Castilla--La Mancha, 13071 Ciudad Real, SPAIN.

\bibitem{Di84}
P. A. M. Dirac, ``The Principles of Quantum Mechanics'',
(Clarendon, Oxford, ed. 4, 1984).

\bibitem{Zu91}
W. H. Zurek, Phys. Today {\bf 44}, No. 10, 36 (1991).
W. H. Zurek, Phys. Rev. D {\bf 24}, 1516 (1981);
{\it ibid.} {\bf 26}, 1862 (1982).
W. G. Unruh and W. H. Zurek, Phys. Rev. D {\bf 40}, 1071
(1989).

\bibitem{Wa85}
D. F. Walls and G. J. Milburn, Phys. Rev. A {\bf 31}, 2403
(1985);
A. O. Caldeira and A. J. Leggett, Phys. Rev. A {\bf 31}, 1059
(1985).

\bibitem{Ne32}
``MathematischeGrundlagen der Quantenmechanik'', J. von Neumann
(Springer, Berlin, 1932).

\bibitem{Wh83}
``Quantum Theory and Measurement'', J. A. Wheeler and W. H.
Zurek, Eds. (Princeton Univ. Press, Princeton, NJ, 1983).

\bibitem{Di95}
D. P. DiVincenzo, Science {\bf 270}, 255 (1995).

\bibitem{Ch95}
I. L. Chuang {\it et al.}, Science
{\bf 270}, 1633 (1995).

\bibitem{Da93}
L. Davidovich {\it et al.}, Phys. Rev. Lett. {\bf 71}, 2360 (1993).

\bibitem{Ga94}
B. R. Garraway and P. L. Knight, Phys. Rev. A
{\bf 50}, 2548 (1994); {\bf 49}, 1266 (1994);
C. C. Gerry and E. E. Hach III, Phys. Lett. A {\bf 174}, 185
(1993).

\bibitem{Go95}
P. Goetsch {\it et al.}, Phys. Rev. A {\bf 51}, 136
(1995).

\bibitem{Ga92}
See, for example, C. W. Gardiner, {\it Quantum Noise},
(Springer--Verlag, Berlin, 1992) and references therein.

\bibitem{Wi87}
D. J. Wineland and W. M. Itano, Physics Today, {\bf 40}, No. 6,
34 (June, 1987).

\bibitem{Ci93}
J. I. Cirac {\it et al.}, Phys. Rev.
Lett. {\bf 70}, 556 (1993); {\bf 70}, 762 (1993).
The way to preparing arbitrary
states of motion is shown in S. A. Gardiner {\it et al} (unpublished).

\bibitem{Po95}
J. F. Poyatos {\it et al.}, Phys. Rev. A (to be published);
S. Wallentowitz and V. Vogel, Phys. Rev. Lett. {\bf 75}, 2932
(1995).

\bibitem{Mo95b}
C. Monroe {\it et al.}, Phys. Rev. Lett. {\bf 75},
4714 (1995); D. M. Meekhof {\it et al} (unpublished).

\bibitem{Mo96}
C. Monroe {\it et al.} (unpublished).

\bibitem{Sc35}
E. Schr\"odinger, Naturwissenschaften {\bf 23}, 807 (1935);
{\bf 23}, 823 (1935); {\bf 23} 844 (1935).

\bibitem{Zo95}
P. Zoller and C. W. Gardiner in ``Quantum fluctuations'', Les
Houches, E. Giacobino {\it et al} (Elservier, NY, in press).

\bibitem{Ma96}
R. L. de Matos Filho and W. Vogel
Phys. Rev. Lett. {\bf 76}, 608 (1996).

\bibitem{Ga86}
C. W. Gardiner, Phys. Rev. Lett. {\bf 56}, 1917 (1986).

\bibitem{Ci95}
J. I. Cirac and P. Zoller,
Phys. Rev. Lett. {\bf 74}, 4091 (1995).

\bibitem{}
We acknowledge discussions with R. Blatt, C. Monroe, D.
Wineland, and W. Zurek. J. F. P. aknowledges a grant
of The Junta de Comunidades de Castilla--La Mancha.
Work supported by Austrian Science
Foundation.


\end{references}
\end{document}